\documentstyle[aps,prb,preprint]{revtex}

\newcommand{\version}{2(June 23 1997) --- printed on \today}

\newcommand{\gsim}{{ _> \atop ^\sim}}

\newcommand{\czn}{\mbox{$\rm Cu_{0.986}Zn_{0.014}GeO_3$}}
\newcommand{\cgo}{\mbox{CuGeO$_{\rm 3}$}}
\newcommand{\zncgo}{\mbox{Cu$_{(1-x)}$Zn$_x$GeO$_{\rm 3}$}}
\newcommand{\znecgo}{\mbox{Cu$_{0.986}$Zn$_{0.014}$GeO$_{\rm 3}$}}
\newcommand{\sicgo}{\mbox{CuGe$_{(1-y)}$Si$_y$O$_{\rm 3}$}}
\newcommand{\tsp}{\mbox{T$_{\rm SP}$}}
\newcommand{\cm}{\mbox{cm$^{\rm -1}$}}

\title{Revival of the spin-Peierls transition in \zncgo\ under pressure}

\author{M. Fischer, P.H.M. van Loosdrecht, P. Lemmens, G. G\"untherodt}
\address{II. Physikalisches Institut, RWTH-Aachen, 
                Templergraben 55, 52056 Aachen, Germany.}

\author{B. B\"uchner, T. Lorenz, M. Breuer}
\address{II. Physikalisches Institut, Universit\"at zu K\"oln,
                Z\"ulpicherstra\ss e 77, 50937 K\"oln, Germany.}

\author{J. Zeman\footnote{on leave from the Institute of Physics, 
Academy of Sciences, Prague, Czech Republic}, G. Martinez}
\address{Grenoble High Magnetic Field Laboratory  MPI/CNRS,
                25 avenue des Martyrs, F38042 Grenoble Cedex 9, France.}

\author{G. Dhalenne, A. Revcolevschi}
\address{Laboratoire de Chimie des Solides, Universit\'{e} de Paris-Sud,
                b\^atiment 414, F-91405 Orsay, France}

\begin{document}
\draft
\maketitle

\begin{abstract}
Pressure and temperature dependent susceptibility and Raman scattering 
experiments on single crystalline \znecgo\   have shown an unusually strong 
increase of the spin-Peierls phase transition temperature upon applying hydrostatic 
pressure. The large positive pressure coefficient  (7.5 K/GPa)
- almost twice as large as for the pure compound (4.5 K/GPa) -
is interpreted as arising due to an increasing magnetic frustration which decreases the 
spin-spin correlation length, and thereby weakens the influence of the non-magnetic 
Zn-substitution.
\end{abstract}


\begin{center}
\footnotesize  Version \version
\end{center}

\section{Introduction}
The spin-Peierls (SP) transition\cite{PYT74} in low dimensional magnetic compounds 
is the magnetic analog of the well known Peierls transition\cite{PEI95} 
in low dimensional metals. 
The interest in this magneto-elastic phenomenon, which has been studied widely in 
organic compounds in the 1980's\cite{BRA83}, 
has been strongly renewed with the discovery of inorganic materials 
(\cgo \cite{HAS93}, $\alpha$-NaV$_2$O$_5$\ \cite{ISO96})
exhibiting this unusual phase transition. 
In particular \cgo\ has attracted strong experimental and  
theoretical 
attention over the past few years\cite{BOU96}. 
There are at least three reasons for this: 
In the first place \cgo\  allows for the growth of large single crystals 
which permits for  
experiments which were difficult to perform in the past. 
Secondly, \cgo\ allows for well controlled substitutions (e.g. \zncgo, \sicgo) 
which provides a useful method to study its structural, 
and in particular its magnetic properties. A surprising  
observation 
is that the SP phase transition is extremely sensitive to substitution 
(5 \% of Zn  substitution for Cu already makes the SP phase disappear). 
Equally surprising is that substitutions, which inherently  
increase the disorder in the system, lead to the formation of a long-range 
ordered antiferromagnetic phase at  
low temperatures\cite{REN95,REG95,HAS95,LEM97}. 

In the third place \cgo\ cannot be considered as a ``classical'' spin-Peierls system. The  
susceptibility in the uniform phase (T$>\tsp$) strongly deviates 
from the Bonner-Fisher behavior \cite{CAS95,RIE95} and
the spin-Peierls transition temperature strongly increases upon applying hydrostatic pressure  
despite an apparently decreasing 
$nn$ exchange interaction\cite{WIN95,TAK95,BUC96}.
There may be several reasons for the deviations from the classical behavior, 
notably a frustrating next nearest neighbor interaction\cite{CAS95,RIE95}, 
the two-dimensionality of the magnetism in the system\cite{UHR97}, 
and the degeneracy of the energy scales for phonon and spin excitations 
in \cgo\ which 
calls for a description of \cgo\  in terms of mixed spin-phonon excitations rather than more 
or less decoupled spin and phonon excitations. At present, however,  
it is still unclear which of the above effects 
plays the dominant  role in \cgo\ , 
although it has been shown that one can certainly not neglect the frustration 
in the system induced by the next nearest neighbor $(nnn)$ interactions. The values reported 
for the frustration in \cgo, based on one-dimensional (1D) adiabatic approximations, range from 
$\alpha=J_{\rm nnn}/J_{\rm nn}=0.24-0.36$\cite{CAS95,RIE95}, 
and may even be higher in \cgo\ under hydrostatic  pressure\cite{BUC96,LOO97}. 
Within the 1D adiabatic approximations one then expects \cgo\ 
to have a spin-gap, even without dimerization, which is small for 
$\alpha_c<\alpha<<0.5$, 
but increases strongly upon approaching the Majumdar-Gosh point ($\alpha=0.5$)\cite{MAJ69}. 

The influence of non-magnetic impurities on one-dimensional spin-Peierls systems has 
been studied 
theoretically by several authors\cite{FUK96,MAR97,FAB97,MOS97}. 
One of the main conclusions drawn from these studies 
is that one can not consider such an impurity as a local site defect. 
In fact, it is expected that 
a non-magnetic impurity induces local antiferromagnetic correlations over a lengthscale 
which is determined by the spin-spin correlation length ($\xi\sim $8-10 sites
\cite{FUK96}). 
This enhancement of the local 
antiferromagnetic correlations may explain the extreme sensitivity 
of the SP phase to substitutions
and possibly also \cite{mosta} the low temperature AF long range order in \zncgo\ and \sicgo. 
Experimentally, evidence for these enhanced correlations may be found in Raman scattering data 
on substituted compounds\cite{LEM97}. The Raman phonon modes activated in the 
SP-phase of pure \cgo\ 
appear  in substituted compounds already at much higher temperatures, and 
are even observed in compounds having a substitution level for which the 
SP-phase has completely vanished. 

Pressure dependent experiments on \cgo \cite{LOO97,GON96}
have shown that \tsp\  strongly increases upon increasing pressure 
($\partial\tsp/\partial p=4.5$ K/GPa), 
in good agreement with earlier predictions derived from zero pressure 
specific heat and thermal expansion experiments \cite{WIN95} using 
the Ehrenfest relation ($\partial\tsp/\partial p=5$ K/GPa). 
It has been argued \cite{BUC96,LOO97} that the pressure dependence of 
\tsp\ may be due to a strong positive pressure coefficient
of the $nnn$ frustration in \cgo\ ($\partial\alpha/\partial p>0$). 
Recently, thermodynamic experiments on substituted compounds
by T. Lorenz {\it et al.} \cite{LOR97A} have led to the 
surprising prediction that $\partial\tsp/\partial p$ of substituted 
compounds is enhanced by a factor of $2-3$ over that of the pure compound.
Motivated by these predictions, we report here on 
the pressure dependence of \tsp\ for
a \znecgo\ single crystal as determined from pressure dependent 
susceptibility and Raman scatterig experiments. 
At low pressure, where the above mentioned prediction should be valid, 
the susceptibility experiments indeed show a strongly enhanced 
$\partial\tsp/\partial p$, quantitatively in agreement 
with the predicted  $\partial\tsp/\partial p\approx 8.1$ K/GPa \cite{LOR97A}. 
In contrast to the pure compound, however, Raman scattering experiments
show that the pressure dependence of \tsp\ strongly deviates from a 
linear behavior, and approaches the pressure dependence of the 
pure compound for higher pressures. The remainder of this paper is organized 
as follows:
Sections II. and III. describe the experimental results 
of the susceptibility and Raman scattering 
experiments, respectively. The subsequent section discusses 
these results in terms of a pressure  
(and thus frustration) dependent spin-spin correlation length. 
Finally, the last section presents the main conclusions of this paper.

\section{susceptibility}

The \czn\  single crystal was grown from the melt by a floating zone method
using an image furnace~\cite{REV69}. For our measurements of the 
susceptibility under pressure we use a piece ($5 \times 3 \times 2$mm$^3$) 
of the crystal studied in Ref.~\cite{LOR97A}.
On the one hand the specific heat and thermal
expansion data presented there prove the high quality of this crystal.
On the other hand, using the same sample enables an unambiguous
comparison between the measured pressure derivatives
of T$_{\rm SP}$ and the predictions from thermodynamic
properties~\cite{LOR97A}.

The ac susceptibility was measured in a pressure range
$0 \le p \le 0.43$\ GPa with a mutual inductance bridge at a frequency of
6.1 KHz in a magnetic field of about 1 Oe. Both,
the primary coil and the astatic secondary coils are
located inside the teflon cup of a piston
cylinder pressure cell which is similar to
that described by J.D. Thompson~\cite{thompson}.
A methanol:ethanol (4:1) mixture
was used as pressure transmission medium
in order to guarantee hydrostatic pressure conditions.
The pressure was controlled at low temperatures by simultaneously
measuring the superconducting transition temperature
of a small piece of lead.

In Fig. 1 the raw data of the ac susceptibility
measurements are presented for several pressures.
The overall temperature dependence of the AC signal
is mainly determined by a smooth background, which
does not change strongly as a function
of pressure. This background signal is nearly temperature independent
above about 12 K and increases at lower temperatures~\cite{frequencyduetopb}.
Due to this background and its pressure dependence
a precise quantitative extraction of the
susceptibility of \czn\ is not possible. In particular,
one can not separate the Curie-like increase of $\chi$
present in doped \cgo\ at low temperatures~\cite{HASE93b} from the
background signal. However,
the pronounced anomalies at the spin-Peierls
transition, which is associated with a rapid decrease of
the susceptibility in a small temperature range,
are clearly visible at all pressures. Besides
the apparent increase of T$_{\rm SP}$ we
do not find significant pressure dependencies
of the susceptibility anomaly at the corresponding T$_{\rm SP}$.

As shown e.g. in Ref. \cite{pouget} the
spin-Peierls transition temperature can be derived precisely
from a sharp maximum of the temperature derivative of the magnetic
susceptibility. The T$_{\rm SP}$ obtained in this way
are shown as a function of pressure in
Fig. 2. Together with the data we show a
line, which represents a pressure derivative of
$\partial T_{\rm SP}/\partial p = 8.1(10)$K/GPa as predicted
from the Ehrenfest relation for $p\rightarrow 0$~\cite{LOR97A}.
The measured, nearly linear,
pressure dependence of \tsp\  up to 0.43 GPa
is in excellent agreement with the predicted
behavior. Fitting the data obtained from the susceptibility
measurements by a linear increase of T$_{SP}$
reveals an only slightly smaller slope of
of $\partial T_{\rm SP}/\partial p = 7.3(3)$K/GPa.
We mention, however, that
this slope systematically increases with decreasing
pressure range considered for the linear fit. For example, fitting
the data up to 0.27 GPa yields 8.2(5)K/GPa,
whereas the slope $\partial T_{\rm SP}/\partial p$
above 0.2 GPa only amounts to 6.3(7)K/GPa.
Thus the susceptibility data are also
consistent with a slight decrease of $\partial T_{\rm SP}/\partial p$
with increasing pressure.
The susceptibility experiments are particularly suited for the low pressure
region ($p<1$ GPa), where also the prediction made in Ref.\cite{LOR97A} 
should hold.
These low pressure data are shown in Fig. 2 (filled symbols) 
which also displays a linear fit to the data (solid line) and the prediction 
made in Ref.\cite{LOR97A} (dashed line). The measured, nearly linear
pressure dependence of $\tsp$ clearly agrees well with the predicted
behavior in this pressure range.

Comparing the data on \czn\  presented so far to the results
for pure \cgo\ it is apparent that the pressure derivative
in the doped compound is significantly larger. In other words,
the reduction of $T_{\rm SP}$ due to doping with 1.4\% Zn
decreases from $\simeq3$K at $p = 0$ to $\simeq 2$K at
$p = 0.5$ GPa. Since an extrapolation of our data
of doped \cgo  would yield the unreasonable result that
$T_{\rm SP}$ increases above the value of pure \cgo\ at high pressures 
($p \gsim 2$GPa), we have studied the behavior in this pressure range
using Raman scattering.

\section{Raman Scattering}

Pressure and temperature dependent Raman scattering experiments have been performed
on 
thin platelets ($100\times100\times40$ $\mu$m$^3$) of the same \znecgo\ single 
crystal as used for the susceptibility measurements. These platelets 
were mounted in a clamp-type diamond anvil cell, 
using a methanol/ethanol mixture as the pressure medium. The cell has 
subsequently been mounted in a He flow cryostat with a temperature regulation 
better than 1K.
Polarized Raman spectra have been recorded 
in a quasi-back\-scat\-tering geometry using the 514-nm line of 
an Ar$^{+}$-laser for excitation and a CCD equipped $\sc DILOR-XY$ 
spectrometer for detection.
The polarization of incident and scattered light have both been chosen 
parallel to the c-axis of the orthorhombic crystal structure ({\it i.e.} 
along the Cu-O chains).

Typical spectra observed for different pressures at $T=7$\ K are presented in
Fig. 3. 
For atmospheric pressure one observes three strong peaks at 
184, 330 and 592 cm$^{-1}$ which have been assigned to A$_g$-modes 
of the orthorhombic crystal structure \cite{DEV94}.
The four additional modes observed at 26, 105, 226, and 370 \cm\ are 
characteristic of the SP phase \cite{KUR94,LOO96}.  
The two modes at 105 and 370 \cm\  can be assigned to Brillouin zone boundary 
phonons, and are activated in the SP-phase through the magnetic and/or structural 
cell doubling.
The low frequency mode at 26 \cm\ arises due to transitions from the singlet 
ground state to the lowest magnetic singlet state\cite{BOU97}. The energy 
of this mode is 
somewhat smaller than that of the corresponding mode in the pure compound (30 \cm),  
indicating that Zn-substituted crystals have a smaller spin-gap. 
This is in good agreement with the observed reduction of the  transition 
temperature in \znecgo\ ($\tsp$[\znecgo]$\approx$ 11 K; $\tsp$[\cgo]$\approx$ 
14 K ). 
Finally, the broad, asymmetric response, peaking at 226 \cm\ is generally 
thought to arise from two-magnon scattering due to the  
Fleury-Loudon \cite{FLE68} mechanism , and should reflect a weighted 
two-magnon density of states.

Similar to the situation in the pure compound\cite{GON96,LOO97}, 
the application of hydrostatic pressure leads to 
drastic changes in the observed Raman spectra. 
With increasing pressure  the two-magnon  peak shifts to lower 
frequencies ( 208 cm$^{-1}$ at 2.9 GPa), reflecting an increasing magnetic frustration in the  
system.
In contrast to this, but for the same reasons, 
the gap-related mode at 26 cm$^{-1}$ rapidly shifts  
to higher frequecies upon applying pressure. Simultaneously this mode also 
broadens, and at 2.9 GPa the singlet response is hardly observed anymore. 
Most likely, this broadening is due to non-hydrostatic effects at higher 
pressure, as is also reflected in the width of the phonon modes.

A good method to determine $\tsp$ for the pure compound is to 
monitor the temperature dependence of the singlet mode\cite{LOO97}.
As it is the case for the triplet spin-Peierls gap\cite{NIS94}, the energy of 
the singlet mode strongly decreases upon increasing temperature until it 
vanishes at $\tsp$. Alternatively, it has been shown that also the intensity 
of the spin-Peierls active 370 \cm\ phonon may be used to determine $\tsp$ 
\cite{GON96}. 
The intensity of this mode shows a temperature behavior characteristic for 
a second order phase transition, {\it i.e.} an intensity which varies as 
the square of the order parameter. This is demonstrated for \cgo\ in Fig.
4a, which shows a comparison between the temperature  
dependence of the intensity of the 370-\cm phonon (filled symbols) with 
that of the the experimentally determined spontaneous strain  
(solid line)\cite{LOR97A}, which should vary as the square
of the structural order parameter. 
For the pure compound the above mentioned
methods yield essentially the same results and are 
in good agreement with the transition
temperature obtained using other methods 
(e.g., thermal expansion, susceptibility,
heat capacity), as shown in Refs. \cite{WIN95,LOR97A,inkommi}.

Since the present data do not allow to use the singlet response as an 
indicator for the presence of the SP phase over the whole pressure range, we 
will use the intensity of the 370 \cm\ mode to determine $\tsp$. 
As an example, the inset of Fig. 4b shows the 
temperature dependence of this mode for $P=1.6$ GPa. 
Relative to the $A_g(2)$\ mode at 347 \cm, this mode, which is observed 
as a relatively strong Lorentzian shaped mode at $T=7$\ K, rapidly looses 
intensity as the temperature increases, 
and finally disappears around $\tsp=18$\ K.
Figure 4b shows the temperature dependence of the intensity 
of the 370-\cm phonon as function of 
temperature (normalized to the intensity at the lowest temperature) 
for $p=0, 1.6,$\ and $2.4$\ GPa. 
For higher pressure ($p>1$\ GPa) the intensity indeed shows the 
expected behavior, {\it i.e.} it 
qualitatively seems to reflect the squared order parameter of a second order 
phase transition. In contrast to this, and to the situation in the pure 
compound, the temperature dependence of the 
intensity of the 370 \cm\ mode at $p=0$ does not seem to reflect 
the square of the structural order  
parameter. 
This becomes more clear from a comparison of the temperature dependence 
of the intensity of the 370-\cm\ mode 
with the spontaneous strain measured on the same sample (Fig. 4b, solid line). 
Although the spontaneous strain still shows the behavior of a relatively well defined second 
order phase transition with $\tsp=11.5$\ K, the intensity of the 
370 \cm\ mode drops much more rapidly with temperature, and, moreover, 
remains non-zero up to temperatures well above the phase transition
\cite{LEMNOTE}. 
These observations put question marks to the interpretation of a structurally 
induced intensity 
for the 370-\cm\ mode in the SP phase. Indeed, the intensity of this mode in 
the pure compound also shows a good correlation with the square of the 
singlet energy (the square of the ``magnetic order parameter'').
This is not surprising since the singlet energy directly correlates with 
the SP gap $\Delta$, which itself is related to the structural order 
parameter $\delta$ ($\Delta=\delta^{2/3}$\ within Cross-Fisher theory 
\cite{CRO79}). This indicates that the intensity of the 370-\cm mode may 
not be induced by the structural distortion in the SP phase, but rather by 
the doubling of the magnetic unit cell \cite{LOO97A}. It furthermore strongly 
suggests that the relation between the structural distortion and the size 
of the SP gap may drastically change upon substitutions, strongly 
deviating from the Cross-Fisher prediction \cite{CRO79}. 

\section{Discussion}

At low pressures the transition temperature in the substituted compound 
indeed increases much faster with pressure 
($\partial\tsp/\partial p\approx 7.5$\ K/GPa) 
as compared to the increase for the pure compound
($\partial\tsp/\partial p\approx 4.5$ K/GPa) (see Fig. 2).
The steep linear behavior observed for low pressures clearly cannot persist 
indefinitely. Indeed, the transition temperatures obtained for higher 
pressure from the Raman data show a pronounced 
non-linear behavior at higher pressures (see Fig. 5, full 
squares). In this regime, the pressure dependence of the transition 
temperature for $\znecgo$ approaches that of the pure \cgo\ compound.

In order to get some understanding of the strong pressure dependence of
$\tsp$\ in \zncgo, consider first the strong dependence of $\tsp$\ on 
substitutions. As has been shown by, e. g., Fukuyama {\it et al.}
\cite{FUK96}, a substitution at a single site does not only perturb 
this site, but leads to the formation of local AF ordering and suppression 
of the dimerization over several lattice sites. The spatial extent of 
the impurity is typically of the order of the spin-spin correlation length 
$\xi\sim(J/ \Delta)a$. This immediately explains the strong reduction
of $\tsp$\ upon substitution, and  may possibly also explain the 
occurrence of a long-range ordered AF ground state in the substituted 
crystals\cite{FUK96,FAB97,MOS97}. As has been pointed out by 
Mostovoy and Khomskii\cite{MOS97}, one expects the SP transition to be 
fully suppressed when the average distance between the impurities 
becomes of the order of the spin-spin correlation length. {\it i.e.} 
for concentrations of about 10 \% ($\xi\sim8 a-10 a$).

The strong increase of $\tsp$\ with increasing pressure in pure $\cgo$ 
has been discussed in terms of a strongly increasing frustration 
\cite{BUC96,LOO97}. This may, of course, also explain part of the 
observed increase of $\tsp$ in the substituted compound. There is, 
however, a second effect which arises from the increasing frustration and is 
of importance in the substituted compounds. 
As the frustration increases, one expects the spin-spin correlation length 
to decrease. Indeed, at $\alpha=0.5$\ (the MG point) one expects the 
ground state to be formed by singlets pairs and to have a correlation length 
of exactly two spins. Numerical Density Matrix Renormalization Group 
calculations of the correlation length have indeed confirmed this picture 
\cite{BOUNOTE}. The consequence of the decreasing correlation length is 
that the influence of impurities is strongly suppressed upon applying
pressure. This actually may 
explain the observed behavior of $\tsp(p)$ in \znecgo. Initially, the 
increase of the transition temperature is driven by a combination of an 
increasing frustration and a decreasing influence of the impurities with
frustration. At higher pressure the effect of the impurities is strongly 
suppressed, and the increase of $\tsp$\ is fully driven by the increasing 
frustration alone. At these pressures one therefore expects only small 
differences between the pure and the substituted compounds, in good 
agreement with the observed behavior.

A further indication of the suppression of the effects of the 
impurities may be found from the temperature dependence of the 
intensity of the 370-\cm mode (see Fig. 4b). 
As mentioned in the previous section, at low pressures the behavior of 
the intensity of this mode does 
not agree well with the temperature dependence of the structural 
order parameter. In addition, the mode remains observable at 
temperatures well above the phase transition. This latter observation
is consistent with the idea that the mode is induced by magnetoelastic 
interactions and the magnetic cell doubling. If indeed the presence of 
impurities leads to enhanced AF correlations around the impurities, 
then this also leads to a local magnetic unit cell doubling, and 
hence to an activation of the 370-\cm\ mode in the Raman spectra. 
Of course, at high enough temperatures the intensity should 
still disappear due to the decreasing correlation length, {\it i.e.} 
the decreasing spatial ``size'' of the impurities. 
At high enough pressure ($p>1$\ GPa) the temperature dependence of the 
intensity of the 370-\cm\ mode is again in qualitative agreement with 
the dependence observed for the pure compound, in good agreement 
with the idea that the influence of the impurities is strongly reduced 
for high pressures.

\section{Conclusions}

In this paper we have shown, by two different methods, that the pressure dependence of the 
spin-Peierls transition temperature in Zn-substituted \cgo\ is strongly enhanced over that 
of the pure compound. The initial-slope data are found to be in excellent agreement 
with the predictions using the Ehrenfest relation\cite{LOR97A}. At higher 
pressure the pressure coefficient $\partial\tsp/\partial p$\ of \znecgo\ approaches that of the pure 
compound. The strong enhancement 
at low pressure, and the approach to the behavior of the pure compound at high pressure
is in good agreement with an increasing frustration in these compounds, which leads to a 
decreasing correlation length and thus to a decrease of the influence of the impurities.

\acknowledgements 
We gratefully acknowledge G. Bouzerar for many fruitful discussions, and 
for the calculations on the frustration dependence of the correlation length.
This work was supported by DFG through SFB 341 and by BMBF 13N6586/8. 
Laboratoire de Chimie des Solids is Unit\'{e} de Recherche associ\'{e}e
CNRS n$^{o}$ 446.

\newpage
\centerline{\bf Figure Captions}

Fig. 1: 
Raw data of the AC susceptibility measurements (output 
signals of the lock-in amplifier) for several pressures. 
The data are subsequently shifted by $10^{-4}$\ V for clarity.\\

Fig. 2: 
Pressure dependence of $\tsp$ as determined from
the susceptibility data of $\czn$ (filled symbols). The 
lines gives the predicted pressure dependencies
from the Ehrenfest relation
(solid line: \znecgo\  $\partial\tsp/\partial p =8.1$ K/GPa (Ref. \cite{LOR97A}).
dashed line: \cgo\ $\partial\tsp/\partial p =5$ K/GPa \cite{WIN95}.)

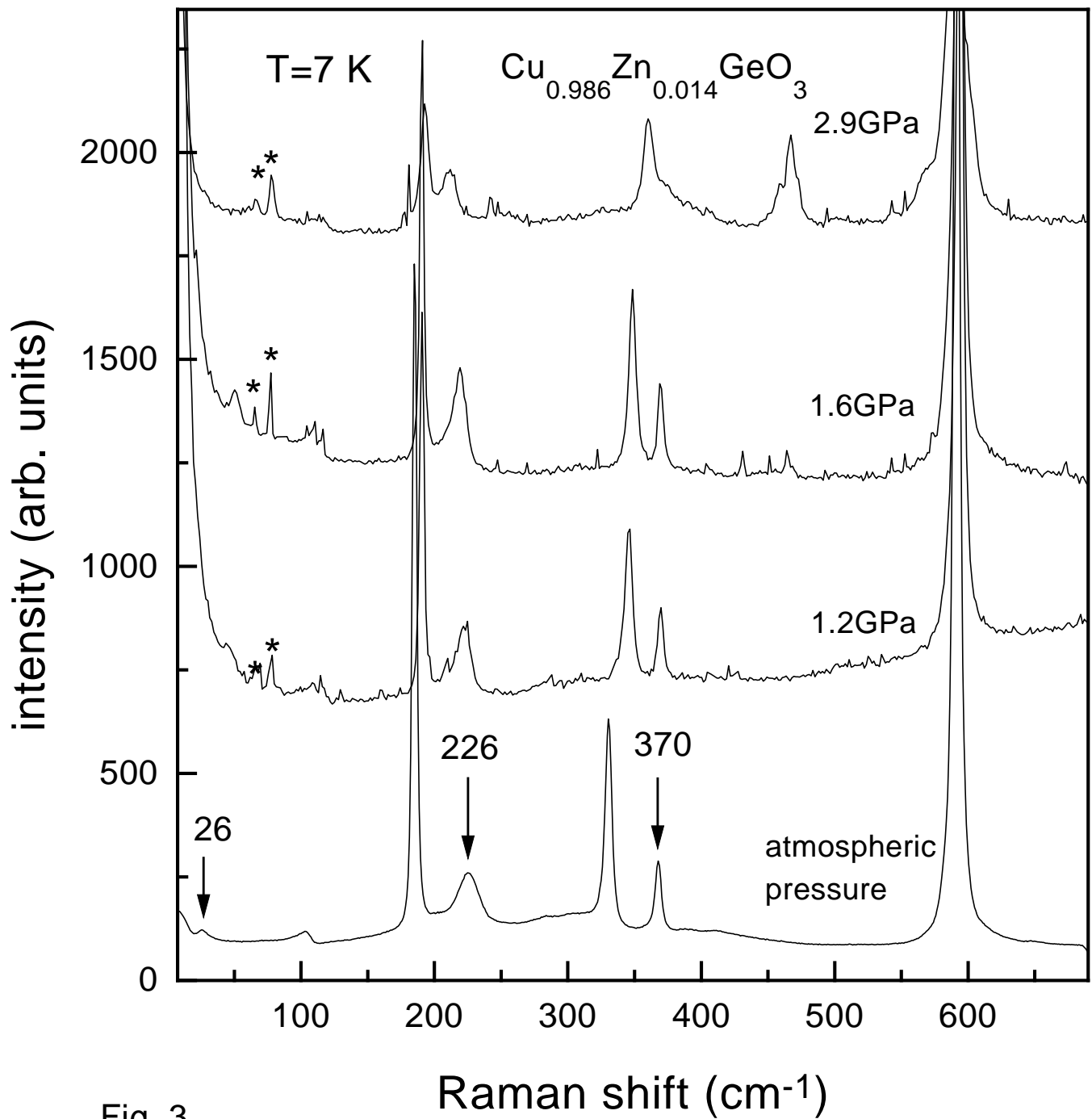
Fig. 3: 
(ZZ) polarized Raman spectra of \znecgo\ 
recorded at T=7 K for $p=0, 1.2, 1.6,$ and $2.9$ GPa.
The curves have been given an offset for clarity. The positions of observable 
Ar$^{+}$-laser plasma lines are denoted by an asterisk.\\

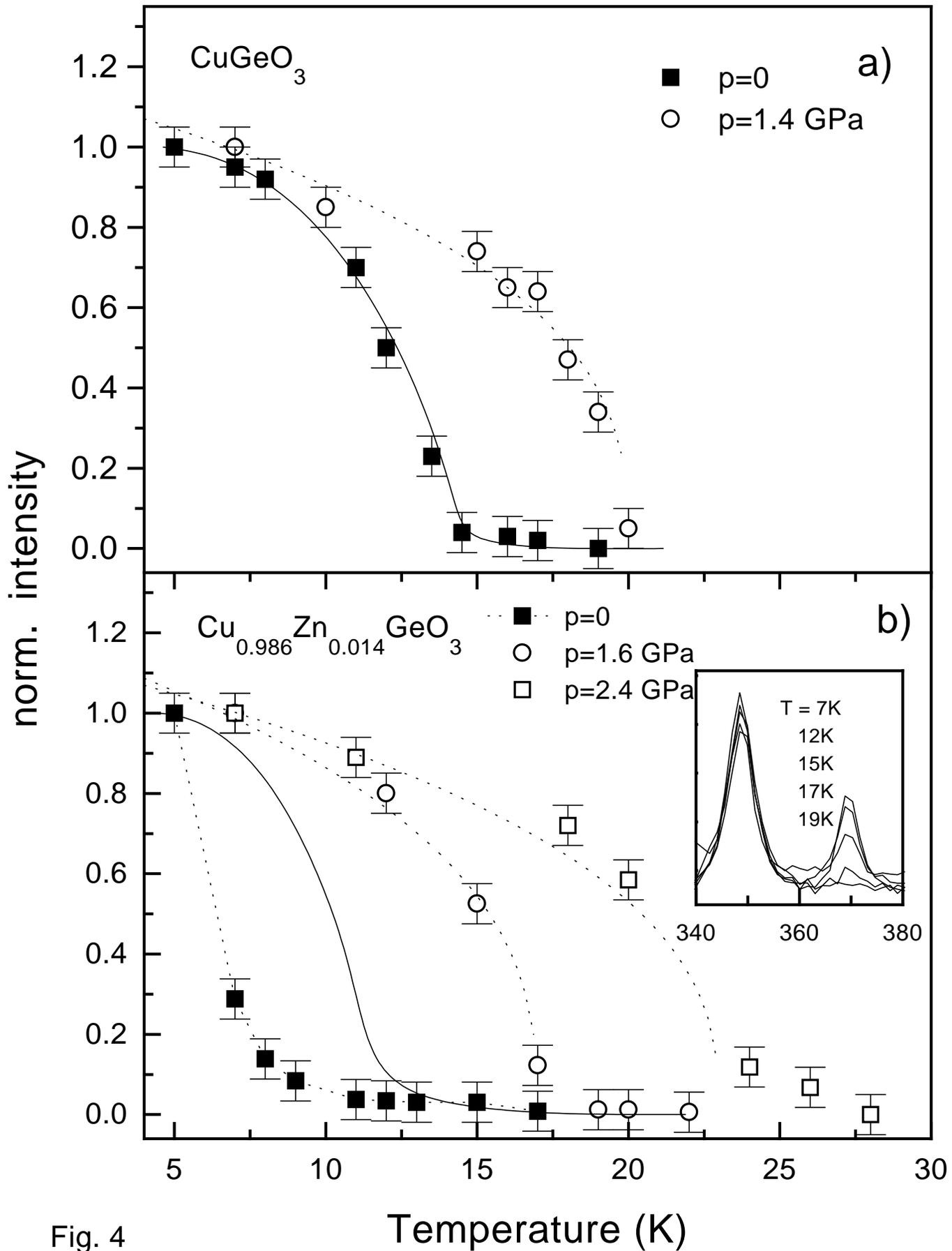
Fig. 4: 
(a) Intensity 370-\cm mode in $\cgo$ as a function of temperature 
for $p=0$ (filled symbols), and  
$1.4$ GPa (open symbols). The solid line gives the experimental 
temperature dependence of 
the spontaneous strain at $p=0$ GPa \cite{LOR97A}. 
Dashes line is a guide to the eye.
(b) Intensity of the 370-\cm\ mode in $\znecgo$ as function of temperature for 
$p=0$ (filled symbols),  
$1.4$ and $2.4$ GPa (open symbols). The solid line gives the experimental 
temperature dependence of the spontaneous strain at 
$p=0$ GPa \cite{LOR97A}. Dashes lines are a guide to the eye.
The inset shows the temperature dependence of the (ZZ) polarized spectrum 
of $\znecgo$ for $p=1.6$ GPa.\\

Fig. 5: Dependence of \tsp\ as function of pressure for \cgo\ (open symbols) 
and \znecgo\ (filled symbols, triangles: obtained from susceptibility data; 
filled squares: 
obtained from Raman data) . The solid lines give the predictions for  
the zero pressure derivatives.


\begin{thebibliography}{prsty}

\bibitem{PYT74}
E. Pytte, Phys. Rev. B {\bf 10},  4637  (1974).

\bibitem{PEI95}
R.~E. Peierls, {\em Quantum theory of solids} (Clarendon, Oxford, 1995).

\bibitem{BRA83}
J.~W. Bray, L.~V. Iterrante, I.~S. Jacobs, and J.~C. Bonner,  in {\em Extended
  linear chain compounds}, edited by J.~S. Miller (Plenum Press, New York,
  1983), Vol.~3.

\bibitem{HAS93}
M. Hase, I. Terasaki, and K. Uchinokura, Phys. Rev. Lett. {\bf 70},  3651
  (1993).

\bibitem{ISO96}
M. Isobe and Y. Ueda, J. Phys. Soc. Jap. {\bf 65},  1178  (1996).

\bibitem{BOU96}
J.~P. Boucher and L.~P. Regnault, J. Phys. I (Paris) {\bf 6},  1939  (1996).

\bibitem{REN95}
J.P. Renard, K. Le Dang, P. Veillet, G. Dhalenne, A. Revcolevschi, and L.P.
Regnault, Europhys. Lett. {\bf 30},  475  (1995).

\bibitem{REG95}
L.P. Regnault, J.P. Renard, G. Dhalenne, and A. Revcolevschi, 
Europhys. Lett. {\bf 32},  579  (1995).

\bibitem{HAS95}
M. Hase, N. Koide, K. Manabe, Y. Sasago, K. Uchinokura, and
A. Sawa, Physica B {\bf 215},  164  (1995).

\bibitem{LEM97}
P. Lemmens, M. Fischer, G. G\"untherodt, C. Gros, P.G.J. van Dongen,
M. Weiden, W. Richter, C. Geibel, and F. Steglich, Phys. Rev. B 
{\bf 55}, 15076 (1997).

\bibitem{CAS95}
G. Castilla, S. Chakravarty, and V.J. Emery, Phys. Rev. Lett. {\bf 75},  1823
  (1995).

\bibitem{RIE95}
J. Riera and A. Dobry, Phys. Rev. B {\bf 51},  16098  (1995).

\bibitem{WIN95}
H. Winkelmann, E. Gamper, B. B\"uchner, M. Braden, A. Revcolevschi,
and G. Dhalenne, Phys. Rev. B {\bf 51},  12884  (1995).

\bibitem{TAK95}
H. Takahashi, N. M\^{o}ri, O. Fujita, J. Akimitsu, and
T. Matsumoto, Sol. St. Comm. {\bf 95},  817  (1995).

\bibitem{BUC96}
B. B{\"u}chner, U. Ammerahl, T. Lorenz, W. Brenig, G. Dhalenne,
and A. Revcolevschi, Phys. Rev. Lett. {\bf 77},  1624  (1996).


\bibitem{UHR97}
G. Uhrig, Phys. Rev. Lett. {\bf 79}, 163 (1997).

\bibitem{LOO97}
P.H.M. van Loosdrecht, J. Zeman, G. Martinez, G. Dhalenne,
A. Revcolevschi, Phys. Rev. Lett. {\bf 79},  487  (1997).

\bibitem{MAJ69}
C.K. Majumdar and D.K. Ghosh, J. Math. Phys. {\bf 10},  1388  (1969).

\bibitem{FUK96}
H. Fukuyama, T. Tanimoto, and M. Saito, J. Phys. Soc. Jpn. {\bf 65},  1182
  (1996).


\bibitem{MAR97}
G. Martins, M. Laukamp, J. Riera, and E. Dagotto, Phys. Rev. Lett. {\bf 78},
  3563  (1997).

\bibitem{FAB97}
M. Fabrizio and R. M\'elin, Phys. Rev. Lett. {\bf 78},  3382  (1997).

\bibitem{MOS97}
M. Mostovoy and D. Khomskii, preprint cond-mat 9703104.

\bibitem{mosta} M. Mostovoy (private communication).
\bibitem{GON96}
A.R. Go{\~n}i, T. Zhou, U. Schwarz, R.K. Kremer, and K. Syassen,
 Phys. Rev. Lett. {\bf 77},  1079  (1996).

\bibitem{LOR97A}
T. Lorenz, E. Gamper, H. Kierspel, S. Kleefisch, B. B\"uchner, A.
Revcolevschi, and G. Dhalenne, Phys. Rev. B {\bf 56}, 501 (1997).

\bibitem{REV69}
A. Revcolevschi and R. Collongues, C.R. Acad. Sci. {\bf 266},  1767  (1969);
A. Revcolevschi and G. Dhalenne, Advanced Materials {\bf 5}, 9657 (1993).

\bibitem{thompson} J.D. Thompson, Rev. Sci. Instrum. {\bf 55}, 231
(1984).

\bibitem{frequencyduetopb} The low temperature increase of the signal
is strongly frequency dependent and does not represent a magnetic
susceptibility.

\bibitem{HASE93b} M. Hase, I.
M. Hase, I. Terasaki, Y. Sasago, K. Uchinokura, and H. Obara,
Phys. Rev. Lett. {\bf 71},  4059  (1993).

\bibitem{pouget} J.P. Pouget, L.P. Regnault, M. Ain, B. Hennion,
J.P. Renard, P. Veillet, G. Dhalenne, and A. Revcolevschi, Phys. 
Rev. Lett. {\bf 73}, 736 (1994).

\bibitem{DEV94}
S.D. Devi{\'c}, M.J. Konstantinovi\'{c}, Z.V. Popovi\'{c}, G. Dhalenne,
and A. Revcolevschi, J. Phys.: Condens. Matter {\bf 6},  L745  (1994).

\bibitem{KUR94}
H. Kuroe, T. Sekine, M. Hase, Y. Sasago, K. Uchinokura,
H. Kojima, I. Tanaka, Y. Shibuya, Phys. Rev. B {\bf 50},  16468  (1994).

\bibitem{LOO96}
P.H.M. van Loosdrecht, J.P. Boucher, G. Martinez, G. 
Dhalenne, and A. Revcolevschi, Phys. Rev. Lett. {\bf 76},  311  (1996).

\bibitem{BOU97}
G. Bouzerar, A.P. Kampf, and F. Sch{\"o}nfeld, preprint cond-mat 9701176.
  
\bibitem{FLE68} 
P. A. Fleury, R. Loudon, Phys. Rev. {\bf 166}, 514 (1968).

\bibitem{NIS94}
M. Nishi, O. Fujita, and J. Akimitsu, Phys. Rev. B {\bf 50},  6508  (1994).

\bibitem{inkommi} T. Lorenz, U. Ammerahl, R. Ziemes,
B. B\"uchner, A. Revcolevschi, and G. Dhalenne, Phys. Rev. B {\bf 54}, 15610
(1997).

\bibitem{LEMNOTE}
Similar observations for Zn- and Si-substituted \cgo\ have been reported in
  ref. \cite{LEM97}.

\bibitem{CRO79}
M.C. Cross and D.S. Fisher, Phys. Rev. B {\bf 19},  402  (1979).

\bibitem{LOO97A}
P.H.M. van Loosdrecht, J.P. Boucher, S. Huant, G. Martinez, G. Dhalenne,
 and A. Revcolevschi, Physica B {\bf 230-232},  1017  (1997).

\bibitem{BOUNOTE}
G. Bouzerar, private communication.




\end{thebibliography}
\end{document}